\documentclass[apj]{emulateapj}

\shorttitle{UV upturn and Environments}
\shortauthors{Yi et al.}

\begin{document}

\title{The UV Upturn in Elliptical Galaxies and Environmental Effects}

\author{Sukyoung K. Yi\altaffilmark{1,3}, Jihye Lee\altaffilmark{1}, Yun-Kyeong Sheen\altaffilmark{1}, Hyunjin Jeong\altaffilmark{2,3}, Hyewon Suh\altaffilmark{1}, and Kyuseok Oh\altaffilmark{1}}
\altaffiltext{1}{Department of Astronomy, Yonsei University, Seoul 120-749, Korea; yi@yonsei.ac.kr} 
\altaffiltext{2}{Korea Astronomy and Space Science Institute, Daejeon 305-348, Korea} 
\altaffiltext{3}{Yonsei University Observatory, Yonsei University, Seoul 120-749, Korea} 


\begin{abstract}

It is suspected that the ultraviolet (UV) upturn phenomenon in elliptical galaxies and extended horizontal-branch stars in globular clusters have a common origin. An extremely high abundance of helium ($Y \sim 0.4$) allows for a working hypothesis, but its origin is unclear. Peng \& Nagai (2009) proposed that primordial helium sedimentation in dark haloes over cosmic timescales may lead to extreme helium abundances in galaxy cluster centers. In this scenario UV upturn should be restricted to brightest cluster galaxies (BCGs) only. This is a clear and testable prediction.
We present tests of this hypothesis using galaxy clusters from Yoon et al. (2008) that were detected by both the Sloan Digital Sky Survey and the Galaxy Evolution Explorer Medium Imaging Survey.
Using a new UV classification scheme based on far-UV, near-UV, and optical photometry 
we found only 5\% of cluster elliptical galaxies show a UV upturn, while 27\% and 68\% are classified
as ``recent star-formation'' and ``UV-weak'' ellipticals, respectively.
The data reveal a modest positive dependence of the UV upturn fraction on galaxy velocity dispersion, which is in agreement with the earlier findings of Burstein et al. (1988) and possibly with the helium sedimentation theory.
However, we do not see any dependency on rank or luminosity of galaxies.
Besides, BCGs do not show any marked difference in UV upturn fraction or strength, which is inconsistent 
with the prediction. We conclude that the aforementioned helium sedimentation theory and its inferred environmental effects are not supported by the available data.

\end{abstract}


\section{Introduction}

The ultraviolet (UV) upturn phenomenon in elliptical galaxies has garnered much attention since its discovery (Code \& Welch 1979; Burstein et al. 1988; see O'Connell 1999 for a review). When the phenomenon was first observed, it was considered a mystery because elliptical galaxies were thought to be almost exclusively composed of old stars. Such stars were believed to be poor UV sources. The situation has significantly changed in recent years. Many elliptical galaxies still seem to form stars at low levels (Yi et al. 2005), and even old stars can be effective UV producers (Greggio and Renzini 1990; Horch et al. 1992; Bressan et al. 1994; Dorman, O'Connell, and Rood 1995; Brown et al. 2000; Buzzoni 2007). Stellar population synthesis models based on the old horizontal-branch star assumption appear to provide the most compelling explanation for UV upturn. However, it is still unclear exactly how some, but not all elliptical galaxies develop such a large number of UV bright stars and exhibit a UV upturn.

Recent research focused on extreme helium enrichment has provided an interesting possibility. In studies on globular clusters, some massive globular clusters, most notably $\omega$~Cen, have been found to exhibit multiple main sequences (Bedin et al. 2004) and interestingly, the bluest main sequence seems to be the most metal-rich (Piotto et al. 2007). This is possible if the most metal-rich stars are extremely helium-rich as well (Norris 2004). Helium is extremely difficult to measure directly and so these claims are dependent upon {\em indirect} deduction based on stellar colors and metal line strengths. An inferred helium abundance of $Y \gtrsim 0.4$ apparently explains the presence of an extended horizontal branch of hot temperatures (Lee et al. 2005). Such UV-bright globular clusters have been found in the elliptical galaxy, Cen A, as well (Sohn et al. 2006; Kaviraj et al. 2007b). This helium abundance value corresponds to $\Delta Y \gtrsim 0.16$ when assuming $Y \approx 0.24$ for non-helium-enhanced stars and poses a serious theoretical challenge with respect to helium-to-metal enrichment scenarios (Choi and Yi 2007; 2008). Internal stellar evolution theory alone does not seem to present a clear solution. Intricate dynamic calculations (D'Ercole et al. 2008; Decressin, Baumgardt, \& Kroupa 2008) have been introduced to provide a working hypothesis.

An alternative solution has been presented by Chuzhoy and Loeb (2004) and Peng and Nagai (2009). Their theoretical models based on primordial helium sedimentation suggest that the environment has a profound effect on the UV spectral evolution of elliptical galaxies. In this work, we focus on the Peng and Nagai study because it has a more dramatic effect. The sedimentation effect is suspected to be larger for more massive galaxies and most extreme for galaxies at the center of galaxy clusters. According to Peng and Nagai (2009), this effect can raise the helium abundance {\em of the gas} in individual galaxies near their center by as much as $\Delta Y_{\rm gal} \sim 0.1$ over a Hubble time, which by itself does not seem large enough to explain the extreme horizontal branch and the UV upturn. 

Sedimentation is thought to be far more dramatic up to $\Delta Y_{\rm clu} \sim 0.4$ towards the galaxy cluster center. This effect is expected to be larger in more massive clusters. According to the theory, enrichment takes roughly 5 Gyr for $\Delta Y_{\rm clu} \sim 0.2$ in massive clusters. If a substantial fraction (e.g., $\sim 30$\%\footnote{Such a large fraction (30\%) of intermediate-age stars in BCGs hypothesized here is generally not supported.} as found in $\omega$~Cen) of stars in the brightest cluster galaxies (BCGs) formed from helium-enriched gas roughly 5 Gyr after the initial starburst in the galaxy, it might reach $\Delta Y_{\rm BCG} = \Delta Y_{\rm gal} + \Delta Y_{\rm clu} \gtrsim 0.2$  and explain the UV upturn.  In addition, the sedimentation effect is supposed to be larger for a more massive cluster halo. This theory allows for a clear and testable prediction: UV upturn should mostly be restricted to BCGs that had substantial residual star formation and is more pronounced in more massive cluster haloes. 

It is unclear whether the UV strengths of globular clusters have any causal connection with the UV upturn phenomenon in elliptical galaxies, as they occupy widely separated parameter spaces.
However, both communities raising the same issue of extreme helium abundance seems to call for an investigation.
We hereby perform a test on the above prediction using Sloan Digital Sky Survey (SDSS) optical and Galaxy Evolution Explorer (GALEX) UV photometric data.

\section{Data}

We have built a database of galaxy clusters that contains both optical and UV data. 
The Yoon et al. (2008) galaxy cluster catalogue provides over-dense regions at $0.05 \leq z \leq 0.1$
in the SDSS Data Release 5 through spatial density measurements. It identifies 924 candidate
clusters and their 8266 member galaxies of all morphological types. 
We use their DR7 photometry data.

GALEX matches have been found in GALEX Medium Deep Imaging Survey (MIS, exposure time
range: 1000--5000s) database (Martin et al. 2005). The GALEX Release 6 photometry has been used. 
Nearby ($z \leq 0.1$) bright elliptical galaxies of  $M_{\rm r} < -20.5$ are supposed to be brighter than the MIS 3-sigma detection limit (Yi et al. 2005; Kaviraj et al. 2007; Schawinski et al. 2007). 
This is so if the galaxies have a virtually flat UV spectrum like M32, which is our base assumption for old
quiescent populations that do not have a large number of UV sources such as young stars or UV upturn stars. 
Regarding the shallower All Sky Imaging Survey (AIS, exposure range: 100--400s) of GALEX, a large fraction of the galaxies are fainter than its limiting magnitudes unless there are ongoing unusual star formation activities.
Thus, the AIS database is inadequate for our investigation. 

The MIS mode detects 76\% of our  $M_{\rm r} < -20.5$ galaxies when 3-sigma near-UV (NUV) detection limits were implemented.
We chose the 3-sigma detection criterion primarily to have a reasonably wide baseline in galaxy mass.

\begin{figure}
\centering
\includegraphics[width=0.45\textwidth]{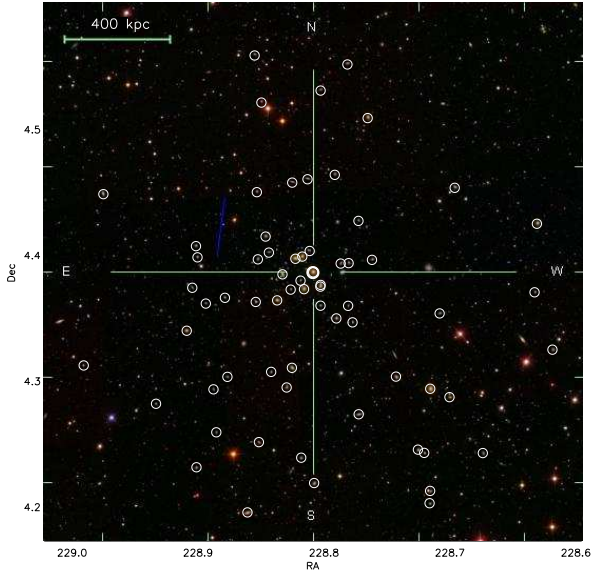}
\caption{
An SDSS image of a representative rich galaxy cluster in our sample, Abell~2048. Smaller circles denote the member galaxies and the thick white circle at the center marks the brightest cluster galaxy. 
}
\label{f1}
\end{figure}

Of the 924 cluster candidates, we use only the richer and so more robust 277 clusters that have at least five member galaxies with $M_{\rm r} < -20.5$. 
This corresponds roughly to half the richness of the Virgo cluster. 
The mass range of these clusters roughly covers from half the Virgo cluster mass to that of Coma.
They contain 2511 galaxies of which 1294 are early type by morphology (visual inspection).
An example cluster is shown in Figure \ref{f1}.

\begin{figure}
\centering
\includegraphics[width=0.5\textwidth]{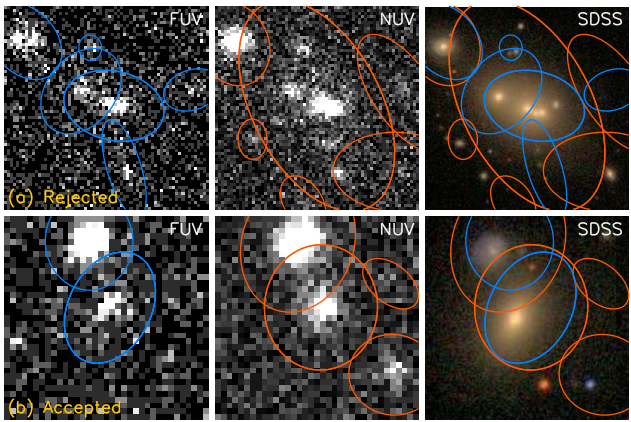}
\caption{
GALEX UV Image inspection. GALEX pipeline apertures for detected objects are marked. The target galaxy is at the center of each panel. The left and middle panels show FUV and NUV images with the aperture for our target galaxy marked. The right panels show the SDSS images with the GALEX apertures of all the GALEX pipeline detections near the target galaxy. The upper/lower panels show a rejected/accepted example.}
\label{f2}
\end{figure}

GALEX data suffer from low spatial resolution and low positional accuracy when compared to optical SDSS data. Thus, source contamination is a serious issue when matching GALEX data to SDSS objects. A small background/foreground UV-bright, but optically-faint object may make correct flux measurements difficult. We visually inspected all GALEX images and their photometric apertures from which magnitudes were derived. Many galaxies have dangerous background or foreground contaminants, but GALEX apertures in the FUV and NUV usually identify them accurately and thus, the photometry data seem reliable. However, in some cases, GALEX photometry fails to identify galaxies or our target is too close to the edge of the GALEX field of view, where photometry is inaccurate. Such galaxies were removed from our sample. Examples of rejected and accepted galaxies are shown in Figure \ref{f2}. Ultimately, 88 galaxies were removed from the sample. We attempted to remeasure their magnitudes manually by rearranging apertures and found that the 88 rejected galaxies are not particularly biased in any way. In other words, they exhibit a color distribution that is similar to an accepted sample. As such, we are not introducing additional bias by removing them from our sample. After removing the 88 galaxies, we have 1206 clean early-type galaxies.

Of the 1206 early-type galaxies, 392 were detected both in far-UV (FUV) and NUV, while 518 galaxies were detected only in NUV; the FUV detector was turned off for many galaxies due to technical issues. 296 galaxies (25\%) were not detected for 3-sigma confidence because they are apparently UV fainter than our base models (M32). This is not surprising because some galaxies are bound to have a fainter UV flux than our base assumption due to various reasons, most notably internal extinction. We took these non-detections into account for the statistical analysis.
We also used Coma cluster data for a local control sample. The Coma cluster is among the most massive clusters known and thus, provides a good test case for analyzing environmental effect.

Galactic foreground extinction has been corrected using maps from the Schlegel, Finkbeiner, \& Davis (1998) and the UV extinctions were estimated via the formulae from Wyder et al. (2005).
All of our photometric data are in the AB mag system.



\section{k correction}

The galaxies in our sample are within a narrow range of redshift ($0.05 \leq z \leq 0.1$), which makes evolutionary correction negligible. We thus consider k-correction only. It is difficult to perform k-correction on UV magnitudes because {\em a priori} knowledge of the entire spectrum should be assumed. We applied simple two-population (young plus old) modeling to the observed photometry, as implemented in previous studies (Ferreras \& Silk 2000; Kaviraj et al. 2007; Jeong et al. 2009; Suh et al. 2010). GALEX FUV to NUV flux ratios are critical to constrain the age of the young population. UV (FUV and NUV) to optical flux ratios mainly constrain the amount of young stars, while optical magnitudes (in our case, SDSS $ugriz$) constrain the age of old components. In this work, we used population synthesis models from Yi (2003).

In the two-component modeling, the old population has a fixed age of 12~Gyr with a luminosity-weighted metallicity of roughly twice Solar (Yi et al. 1999). The young component has the solar metallicity and is allowed to vary in age ($0.01 \le t_{young} \le $10~Gyr) and mass fraction ($10^{-4} \le f_{young} \le 1$). To constrain these two parameters, we fit the models to the observed colors and compute the associated $\chi^{2}$ statistic. The choice of the metallicity for young stars is arbitrary. One might think that younger stars achieve a higher metallicity than the bulk stellar population through chemical recycling. However, the recent work of Crockett et al. (2011) reports that the young stars in residual star formation (RSF) ellipticals have a lower metallicity probably due to the infall of unprocessed gas from outside. Considering the complexity, we arbitrarily assume the solar metallicity for young stars.  For more details, readers are referred to the work of Jeong et al. (2009).

\begin{figure}
\centering
\includegraphics[width=0.45\textwidth]{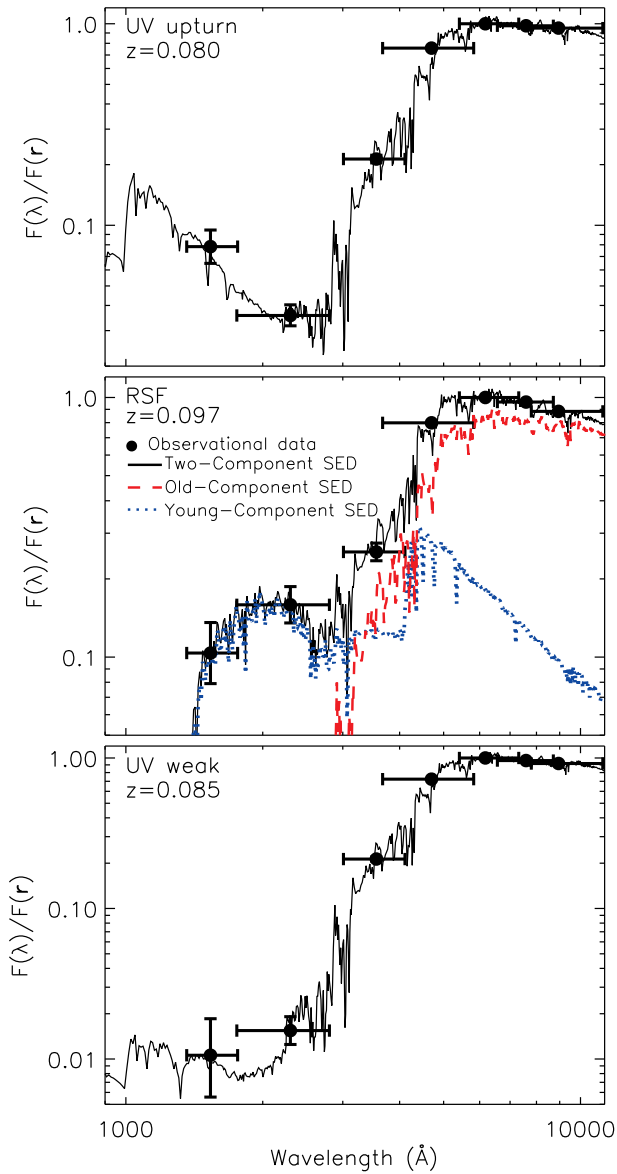}
\caption{
Spectral fits for the estimation of population parameters and $k$ corrections. 
Top panel: An example of UV upturn galaxies. GALEX $FUV$, $NUV$, and SDSS $ugriz$
photometry are shown with band widths (horizontal) and measurement errors (vertical). The spectral fit is based on the Yi (2003) population models.
Middle panel: An RSF elliptical case. The fit employs two (old and young) component populations
to match the data. Bottom panel: A UV-weak case, the majority of elliptical galaxies.
}
\label{f3}
\end{figure}

Representative continuum fits to the observed photometry are shown in Figure \ref{f3}. In the top panel, a galaxy that is closely matched by a UV upturn galaxy model is displayed. The model that fits this galaxy consists purely of old and metal-rich stars. The middle panel in Figure \ref{f3} shows a galaxy that is best fit by a composition of a dominant (99.9\% in mass) old ($\sim 12$~Gyr) population and a tiny (0.1\%) young (0.36~Gyr) population. While we find the best fit model through $\chi^2$ minimization, age-metallicity degeneracy prevents us from determining the values of the metallicities of the two populations. In this sense, our age estimates are model dependent. However, continuum fits are of high statistical significance and thus, k-correction based on the fits bears only a small uncertainty.

Once we find the best fits to the photometry we use the fitting model spectra to derive
k-corrections for individual galaxies. The typical (mean) values of k-correction for our galaxies
are (0.12, 0.23, 0.31, 0.23, 0.08 0.05, 0.06) mag for ($FUV$, $NUV$, $u$, $g$, $r$, $i$, $z$) bands, respectively.  The impact of k-correction will be discussed later in Section 5.2.1.

\begin{figure}
\centering
\includegraphics[width=0.5\textwidth]{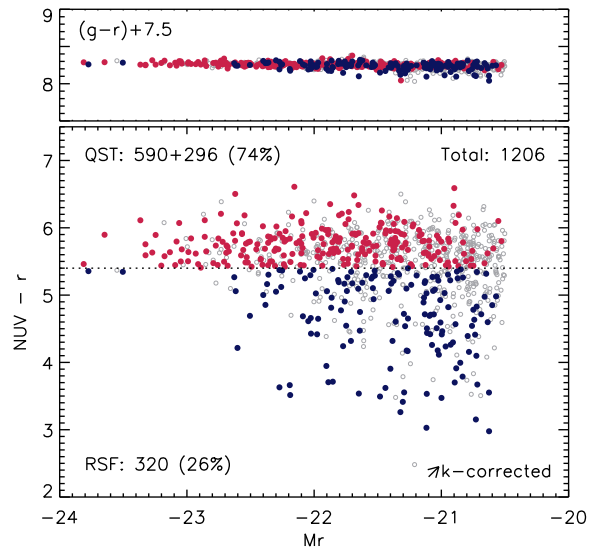}
\caption{
Color-magnitude diagrams of cluster elliptical galaxies (k-corrected). 
The top panel shows a tight optical relation and the bottom panel shows a UV-optical relation with 
a large spread. The typical k-correction vector is presented by an arrow. The empirical range
of ``quiescent'' elliptical galaxies (red dots), $NUV$$-$$r > 5.4$, has been adopted from Yi et al. (2005).
Blue symbols mark the residual star formation (RSF) galaxies containing young stars.
Grey circles denote non-detections in the FUV.
}
\label{f4}
\end{figure}

Figure \ref{f4} shows a color-magnitude diagram for our cluster elliptical galaxies.
Their optical diagram is very tight as has been known for many years, but their UV-optical
diagram shows a large spread as found only recently (Yi et al. 2005).
Elliptical galaxies in galaxy clusters appear to be dominantly, but not exclusively, passive.

\section{UV Classification Scheme}

We attempt to classify cluster elliptical galaxies by UV-optical spectral morphology.
We can use the results of the spectral fits described in the previous section to classify galaxies.
However, they are model dependent. So we take a simpler and empirical route. 

Our new classification makes use of three color criteria.
The first is the $NUV$$-$$r$ color. 
Yi et al. (2005) assumes that the NUV flux of NGC~4552, a famous UV upturn galaxy,
is an empirical upper limit that can be produced by an old stellar population.
This corresponds to $NUV$$-$$r = 5.4$.
This is of no doubt debatable.
NGC~4552 is indeed among the UV strong passive galaxies in the local universe, but
there can be other passive galaxies that have stronger NUV flux.
On the other hand, even UV-red ($NUV$$-$$r > 5.4$) elliptical galaxies may still contain a low level of RSF.
In fact, there are galaxies with such red colors but with signs of young stars (Jeong et al. 2009).
Despite the criticisms we still find it useful to have a hypothetical criterion for
``quiescent'' vs RSF elliptical galaxies. We adopt $NUV$$-$$r = 5.4$, represented by the horizontal dotted line in the lower panel of Figure \ref{f4} and the vertical dotted line in Figure \ref{f5}.

The second is the $FUV$$-$$NUV$ color. 
The UV spectral slope is indicative of the temperature of hot stars. 
$FUV$$-$$NUV =0.9$ corresponds to a flat UV spectrum in the $\lambda$$-$$F_\lambda$ 
domain as shown in Figure \ref{f3}. Hence, $FUV$$-$$NUV <0.9$ is used as a criterion for 
a rising slope in the UV with decreasing wavelength, a necessary condition for UV upturn.
This is the horizontal dotted line in Figure \ref{f5}.

The last is the $FUV$$-$optical color. 
To qualify as a UV upturn galaxy it must have a high relative flux of UV to optical.
The typical value of $FUV$$-$$V$ for the quiescent $FUV$$-$$NUV =0.9$ galaxies 
is roughly 6.21 (Bureau et al. 2011), and with a band correction between $V$ and $r$
our criterion for the UV upturn becomes $FUV$$-$$r < 6.6$.
This is shown in Figure \ref{f5} as a slanted dashed line.
The classification scheme is summarized in Table 1.

\begin{table}
 \centering
  \caption{UV upturn criteria}
  \begin{tabular}{l l}
  \tableline \tableline
    \multicolumn{1}{c}{Criterion} & \multicolumn{1}{c}{Reason} \\
  \tableline
     $NUV$$-$$r > 5.4$   		&  Devoid of young UV-bright stars \\
     $FUV$$-$$NUV < 0.9$   	&  UV rising slope with decreasing $\lambda$ \\
     $FUV$$-$$r < 6.6$   		&  FUV flux pronounced \\
 \tableline \tableline
\end{tabular} \label{tbl:sampling_crit}
\end{table}

\section{Results}

\subsection{Coma cluster}

\begin{figure}
\centering
\includegraphics[width=0.5\textwidth]{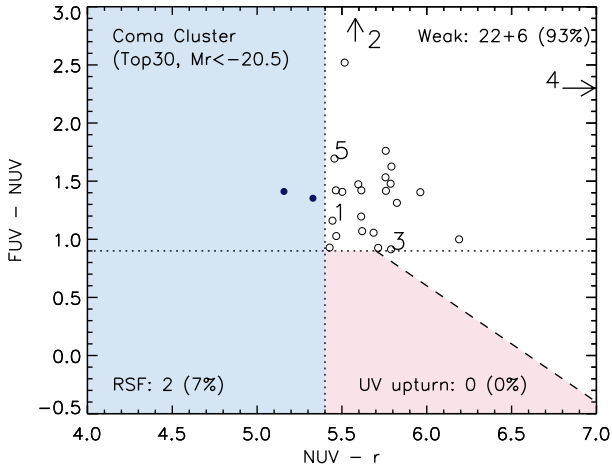}
\caption{Two-color diagram for UV upturn classification for the 30 brightest elliptical galaxies in the Coma cluster.
The vertical dotted line marks the criterion for young stars, the horizontal dotted line denotes a rising UV slope, and the slanted dashed line for the UV strength required to be classified as a UV upturn galaxy. RSF stands for residual star formation. The five brightest early-type galaxies are marked by their rank. None of them is classified as a UV upturn galaxy.
}
\label{f5}
\end{figure}

Classification was first performed on Coma (Figure \ref{f5}), a cluster famous for its size and richness in elliptical galaxies. While  its effect is small due to its low redshift, k-correction was performed. Of the 30 brightest elliptical galaxies, none 
was classified as UV upturn galaxies, two (7\%) as RSF galaxies, and 28 (93\%, including 6 non-detections) as UV-weak galaxies. The BCG (NGC~4889) of Coma is not classified as a UV upturn galaxy. In fact, none of the five brightest elliptical galaxies in Coma is classified as UV upturn galaxies. 
Only 2 RSF galaxies were found in Coma, making it one of the quietest clusters in the local universe. 

\subsection{SDSS-GALEX clusters}

\begin{figure}
\centering
\includegraphics[width=0.5\textwidth]{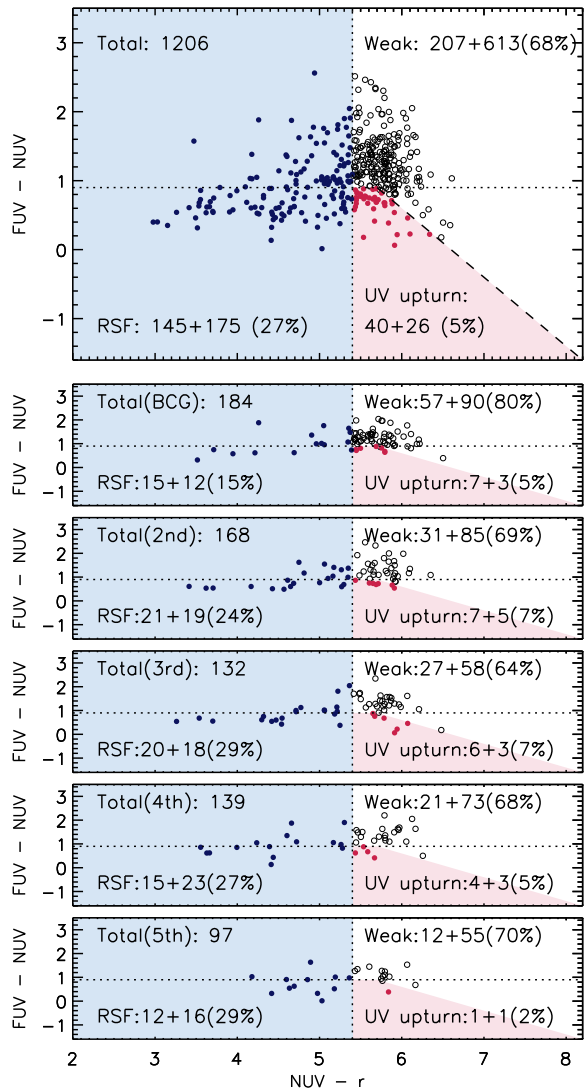}
\caption{Same as Figure 5 but for SDSS-GALEX cluster galaxies.
The top panel shows the whole galaxies with FUV and NUV detections.
For each class, the number of detected galaxies (e.g., 40 in case of UV upturn) and the number of simulated galaxies from non-detections (e.g., 26 in case of UV upturn) are used to derive the fraction of each class.   
From the second to bottom panels, the first (i.e. BCG) through fifth brightest elliptical galaxies are presented, respectively. Simulations for non-detections have been performed in all panels. In each panel, the first integer shows the number of galaxies with both FUV and NUV detections, and the second integer shows the number of galaxies without FUV detections derived through statistical simulations (see text).
}
\label{f6}
\end{figure}

We apply the same UV classification scheme to the SDSS-GALEX overlap sample of
galaxy clusters: 392 cluster early-type galaxies with both FUV and NUV detections.
Figure \ref{f6} shows the whole sample in the top panel, and the first (i.e. BCG) through fifth brightest elliptical galaxies in clusters in the second through the bottom panels, respectively. We call this brightness rank ``rank''.
Some clusters have spiral galaxies in the top five brightest galaxy list, but
spirals are not included in our analysis. So, the number of elliptical galaxies is different for different ranks.
When only galaxies with both FUV and NUV detections are counted, 
10\% (40/392), 37\% (145/392), 53\% (207/392) are classified as UV upturn, RSF, and
UV-weak galaxies, with k-correction. 

We then consider the effect of non-detections.
In our sample, 296 galaxies were not detected in any UV band simply because they are too faint.
We classify them as UV-weak galaxies.
On the other hand, 67\% (814/1206) of the galaxies were not detected in the FUV mainly because the FUV detector was off.
Without their FUV measurements it is not possible to use our classification scheme directly, and so we
perform a statistical simulation.
Of the FUV non-detections, 64\% (518/814) were detected in the NUV. We can use their $NUV$$-$$r$ colors to see whether they are RSF or ``quiescent'' (including UV upturn and UV-weak) galaxies first.
We identified 175 RSF and 343 quiescent ellipticals this way.
The 343 ``quiescent'' ellipticals are divided  into UV upturn or UV-weak categories following
the fractional distributions found from the galaxies with both FUV and NUV detections (10\%:53\%).
Then, we re-derive the fractions including these non-detections.
We iterate this process until the fractions become stabilized to under 1\% level.  
After the iterations, 26 and 317 are classified as UV upturn and UV weak, respectively.
The final ratios are 5, 27, 68\% for UV upturn, RSF, and UV-weak classes, respectively\footnote{It is evident in Figure 6 that the exact classification fractions depend on the accuracy of classification criteria.}.
Compared with Coma in the previous section, both the UV upturn  and RSF fractions are higher.
The UV upturn fraction seems too low for helium sedimentation to have occurred in all cluster halo potentials.
This will be discussed later.

We do not notice any difference in the UV {\em strength} between BCGs and others, either.
The locations of UV upturn galaxies in different panels of Figure \ref{f6} do not reveal any systematic difference. For example, the UV strength (in $FUV$$-$$NUV$ and $FUV$$-$$r$) does not show any rank dependence.
In fact, BCGs show relatively lower UV upturn strength.

\subsubsection{UV upturn parameters}

\begin{figure}
\centering
\includegraphics[width=0.5\textwidth]{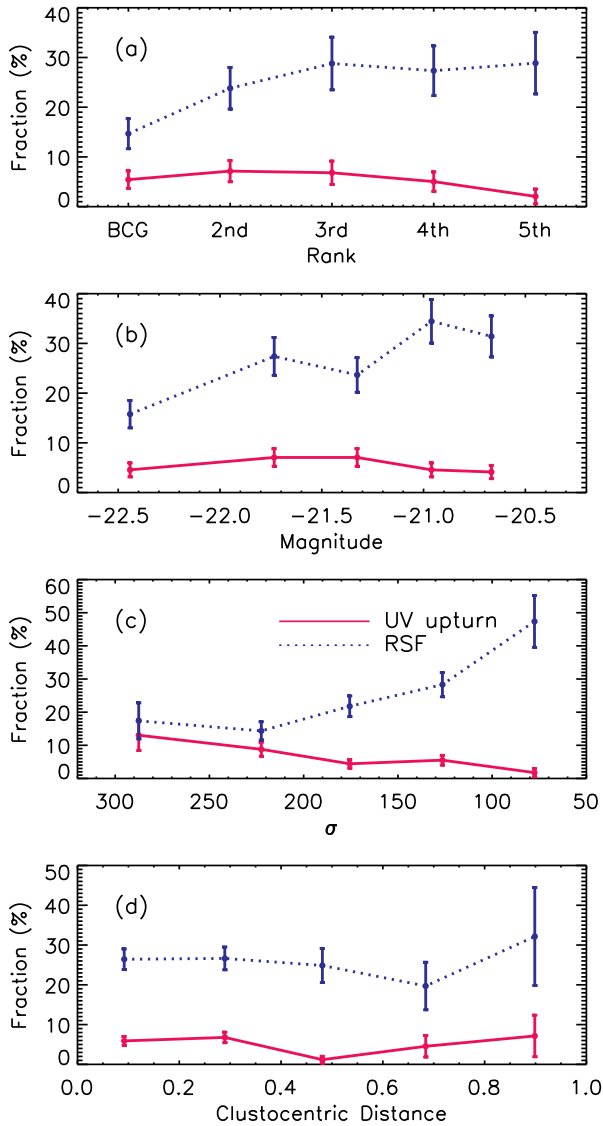}
\caption{UV upturn and RSF galaxy fractions with respect to rank (a), magnitude (b), galaxy velocity dispersion (c) and clustocentric distance (d). Rank refers to the brightness rank, and the clustocentric distance bins are normalized to each cluster virial radius. K-correction has been performed.
}
\label{f7}
\end{figure}

\begin{figure}
\centering
\includegraphics[width=0.5\textwidth]{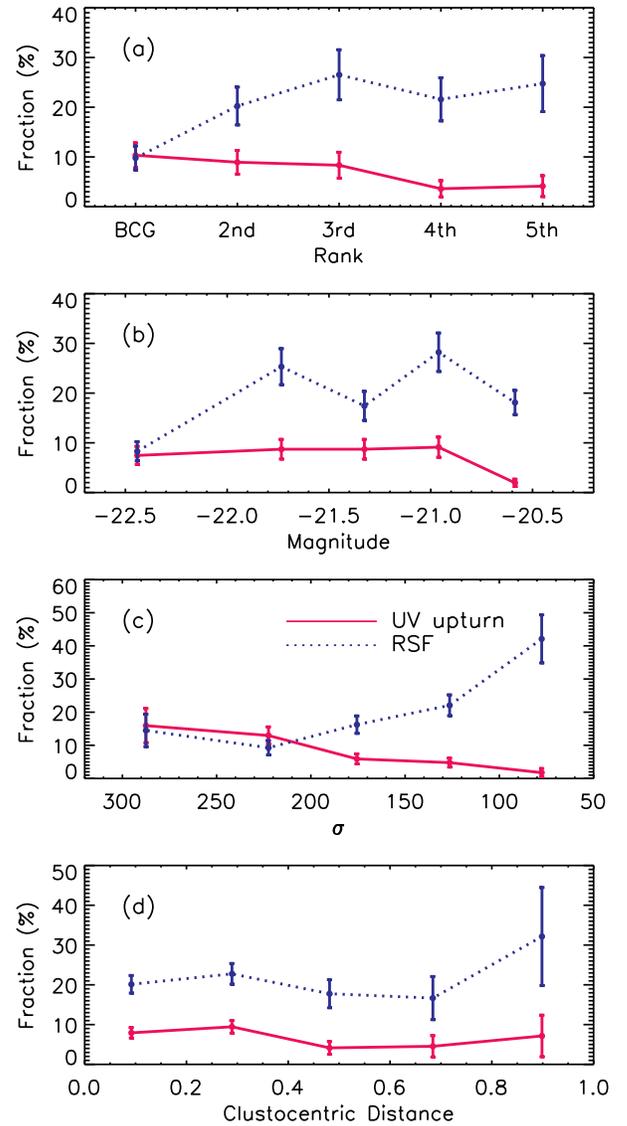}
\caption{Same as Figure \ref{f7} but without k-correction.
}
\label{f8}
\end{figure}

Figure \ref{f7} presents the fractions of UV upturn and RSF elliptical galaxies with respect to
the brightness rank, brightness (in magnitudes), galaxy velocity dispersion and clustocentric distance.
The top panel shows the rank dependence. The key point in this investigation is to check whether
BCGs are so significantly different in the UV upturn strength as the Peng \& Nagai theory predicts.
However, BCGs do not seem to show any particular enhancement in the UV upturn fraction, as was the case in Coma. We found no trend with rank or magnitude. BCGs do not appear special.
On the other hand, we found a clear dependence of UV upturn fraction on velocity dispersion.
Such a mass dependence has been noticed earlier by Faber (1983), Burstein et al. (1988) and 
recently by Carter et al. (2011).
It might be viewed compatible with the helium sedimentation scenario on individual galaxies {\em excluding} cluster halo sedimentation.
The bottom panel of Figure \ref{f7} shows that the central position in the cluster does not appear special, either.

K-correction is particularly important when we deal with spectral ranges with a steep slope.
When we do {\em not} apply k-correction to our observed data, we have Figure \ref{f8} instead of Figure \ref{f7}.
Our data show that BCGs and second-rank ellipticals are more confined to the border regions between UV upturn and UV-weak domains in the classification scheme than fainter galaxies (see the second panel in Figure \ref{f6}) and thus are more vulnerable to the details on k-correction.
Applying k-correction changes the total UV upturn fraction from 8\% to 5\% and the total RSF fraction from 22\% to 27\%.
UV upturn fractions are still very low and BCGs do not appear special considering the underlying mass dependence.

\begin{figure}[h]
\centering
\includegraphics[width=0.5\textwidth]{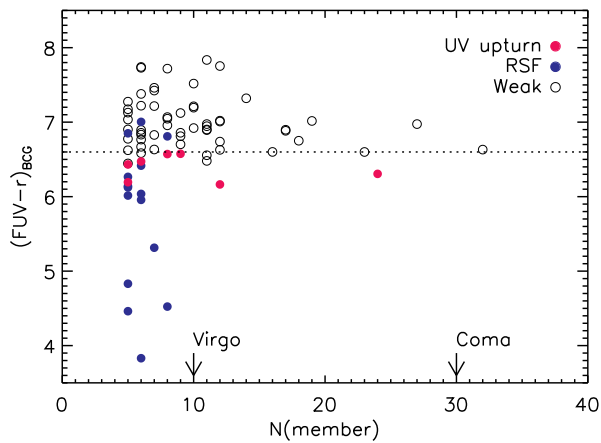}
\caption{The FUV to optical colors of BCGs with respect to cluster mass (number of member galaxies of $M_{\rm r} < -20.5$). The Virgo and Coma clusters are marked.}
\label{f9}
\end{figure}

Figure \ref{f9} shows the FUV to optical colors of BCGs with respect to cluster size, i.e., the number of cluster member galaxies of $M_{\rm r} < -20.5$ in this case. Red symbols indicate the BCGs classified as UV upturn galaxies. The Virgo and Coma clusters are also indicated. The UV strengths of the BCGs do not seem to be correlated with the size of the cluster potential well. Loubser \& S\'{a}nchez-Bl\'{a}zquez (2010) found the same result from their study on the UV strength of BCGs and host cluster properties.

Another test has been performed regarding the position of BCGs in cluster halo potentials. When we inspect galaxy clusters BCGs are not always at the apparent cluster center. In fact, the two best known local clusters, Virgo and Coma, have two seemingly-competing giant ellipticals near the cluster center, but it is less than obvious which one is the true BCG. If helium sedimentation is concentrated on the cluster dark halo center rather than on the baryon center (i.e., BCG) and if BCGs are dislocated from the dark halo center in some clusters for some reason (e.g., not fully virialized), these BCGs would not feel the benefit from the sedimentation. So, we generated a subsample of the rich relaxed clusters. We mean $N_{\rm member} > 10$ with bright ($M_{\rm r} < -20.5$) galaxies by ``rich'' and BCGs being near the apparent center of the clusters by ``relaxed''. We could not notice any significant difference in the UV upturn strength or fraction between our whole cluster sample and this extra-rich cluster sample, however.

Of the 392 elliptical galaxies with FUV and NUV detections, only 6 (2\%) are classified as active galactic nuclei (AGN) when the Baldwin, Philips, and Terlevich (1981) diagnostics are applied on the SDSS spectra.
Thus, the presence of AGN in the sample hardly affects the results.
Different choices of RSF criterion (e.g., $NUV$$-$$r < 5.0$) that are reasonable do not affect our results much, either.

\subsubsection{RSF parameters}

Figure \ref{f7}-(a) shows that BCGs are less likely to have recent star formation.
BCGs likely accrete more gas from the cluster environment than smaller galaxies but also
heat them up through various kinetic and AGN-related processes, making them hostile
environment for residual star formation.
RSF fractions are sensitive to brightness (and rank) and especially on velocity dispersion.
This has been noticed earlier and extensively discussed (e.g., Schawinski et al. 2006).
We detect no clear RSF trend on clustocentric distance.

\section{Discussion}

In the Peng \& Nagai scenario, the UV upturn is the result of three essential elements:
(1) the assumption that extreme-helium stars evolve into UV-bright phases, (2) helium sedimentation, and (3) prolonged star formation.
Discussing the first condition is beyond the scope of this paper anyway.
Considering the amount of helium sedimentation alone, BCGs become the best candidate to show a UV upturn.
The last element can be examined from the results of other studies.
 
Our test is based on a sample that is an order of magnitude larger than what was previously available. We also used a new {\em empirical} classification scheme to identify UV upturn, RSF, and UV-weak galaxies. 

In cluster environments, roughly 5\% of elliptical galaxies are classified as UV upturn galaxies.
Brighter galaxies appear to show a UV upturn more frequently, in agreement with an earlier report by Burstein et al. (1988), but the mass (in rank and brightness) dependence is weak at least for the narrow
mass baseline of our sample.
BCGs show a UV upturn slightly more frequently than fainter galaxies do, but this does not seem significant in any way as predicted by the Peng \& Nagai scenario, especially when considering the underlying mass dependence.

If helium sedimentation did indeed occur over the Hubble timescale in all large haloes as predicted, 
this fraction (5\%) appears to be too small.
Such findings may mean that the majority of cluster elliptical galaxies,  especially BCGs, did 
not have an extended star formation history hence did not benefit from helium sedimentation.
This is compatible with the general understanding on cluster elliptical galaxy evolution (e.g., 
Bower, Lucy, \& Ellis 1992).
It is then suggested that UV upturn galaxies are rare ellipticals with an extended star formation history.
This is however in contrast to empirical findings.
Burstein et al. (1988) found that UV stronger galaxies tend to be more abundant in magnesium, a representative alpha element. 
Such a finding was later confirmed by Bureau et al. (2011).
As alpha enhancement is generally considered to be the result of a short starburst on the order 1~Gyr or less
(e.g., Worthey, Faber, \& Gonzalez 1992), such an enhancement seems to contradict the requirement of this particular hypothesis.

If helium sedimentation indeed took place as theory suggests, there might have been reasons for that BCGs do not appear special in the UV.
First, some non-BCG satellite (but massive) ellipticals may have been BCGs in the previous cluster halo before halo mergers. For example, one can assume that more than a couple of cluster halo mergers occurred in the last few billion years to make one present-day cluster (which is probable).
Assume that BCGs in the merging clusters are still orbiting as satellites in the merger remnant cluster.
The lifetimes of the satellites that were formerly BCGs depend on the details of the merger conditions such as halo mass ratios and orbital parameters (Boylan-Kolchin, Ma \& Quataert 2008; Kimm, Yi \& Khochfar 2011),
but they can be on the order of a few billion years. 
If so, the difference between the
present-day BCGs and massive satellite ellipticals may not be as large as the closed-box assumption taken in our investigation. 
In this regard, one may consider all the bright satellite ellipticals in a cluster ``former BCGs''.
Such ellipticals may have then benefitted from the helium sedimentation from their former halo environments
and so exhibit a modest UV upturn.
However, the Burstein et al. relation between magnesium abundance and UV strength still contradicts such a notion. Besides, the ``former BCG'' argument seems a bit too contrived. 


We stress that our study takes a purely empirical approach. But this raises an issue regarding the time evolution of UV upturn. The mean redshift of our sample is roughly 0.07, which corresponds to a lookback-time of 0.9\,Gyr from the local universe. Some theories of UV upturn (e.g., Yi et al. 1999; Buzzoni 2007) predict a fast evolution of UV strength with time, while different predictions are available too (e.g. Han et al. 2007). The fast-evolution models suggest that the mean UV strength may have been slightly lower at $z=0.07$ than today. We admit that our analysis could be subject to such details. But we believe that the difference in lookback-time in this case is not large enough to cause a major threat to the interpretation. Besides, while the redshift evolution might be real, we still do not see any environmental effect in our sample (at $z=0.07$) in the manner that has been predicted by the helium sedimentation theory.  

We conclude that the our data are not compatible with the current version of  the helium sedimentation theory. 
Environmental effects on the UV strength of elliptical galaxies, if any, seem elusive. 

\section*{Acknowledgments}
We thank the anonymous referee for several clarifying questions including the classification scheme and the time-evolution of UV upturn in population models. We also thank Daisuke Nagai for his open-minded response to our draft.
SKY acknowledges support from the National Research Foundation of Korea to the Center for Galaxy 
Evolution Research, Korea Research Foundation Grant (KRF-C00156), Doyak grant (No. 20090078756), and from Yonsei University Observatory.
This project made use of the SDSS optical data, GALEX ultraviolet data, and the NASA/IPAC Extragalactic Database.

\clearpage

\end{document}